\documentclass[aps,pra,twocolumn,amsmath,amssymb,showpacs,showkeys]{revtex4}
\usepackage{graphicx}% Include figure files
\usepackage{dcolumn}% Align table columns on decimal point
\usepackage{bm}% bold math

\def\LP{\left(}
\def\RP{\right)}

\def\LB{\left\{}
\def\RB{\right\}}

\newcommand{\eq}[1]{Eq.~(\ref{#1})}    %% Eq. (...)
\begin{document}
\title{Entangled three-particle states in magnetic field: Periodic correlations and density matrices}

\author{Amitabha Chakrabarti}
  \email{chakra[at] cpht.polytechnique.fr}
  %\homepage{http://fiquant.mas.ecp.fr/chakraboa}
  \affiliation{Centre de Physique Th\'eorique, \'{E}cole Polytechnique, 91128 Palaiseau Cedex, France}

\author{Anirban Chakraborti}
  \email{anirban.chakraborti [at] ecp.fr}
  %\homepage{http://fiquant.mas.ecp.fr/chakraboa}
  \affiliation{Laboratoire de Math\'{e}matiques Appliqu\'{e}es aux Syst\`{e}mes, \'{E}cole Centrale Paris, 92290 Ch\^{a}tenay-Malabry, France}

\date{\today}

\begin{abstract}
We present a novel study of the time evolutions of entangled states of three spin-1/2 particles in the presence of a constant external magnetic field, which causes the individual spins to precess and leads to remarkable periodicities in the correlations and density matrices. The emerging patterns of periodicity are studied explicitly for different entangled states and in detail for a particular initial configuration of the velocities. Contributions to precession of anomalous magnetic moments are analysed and general results are also obtained. We then introduce an electric field orthogonal to the magnetic field, linking to the preceding case via a suitable Lorentz transformation, and obtain the corresponding Wigner rotations of the spin states. Finally, we point out for the first time that the entangled states corresponding to well-known ones in the study of 3-particle entanglements, may be classified systematically using a particular coupling of three angular momenta.
\end{abstract}



\maketitle

\section{Introduction}

Schr\"{o}dinger had pointed out long time ago \cite{intro1} that quantum entanglement is a crucial element of quantum mechanics. In fact quantum entanglement is one of the most peculiar feature that distinguishes quantum physics from classical physics and lies at the heart of what is now quantum information theory. Quantum physics allows correlations between spatially separated systems that are fundamentally different from classical correlations, and this difference becomes evident when entangled states violate Bell-type inequalities that place an upper bound on the correlations compatible with local hidden variable (or local realistic) theories \cite{intro2}. In the last two decades research has been very focused on quantum entanglement because the field of quantum information theory (cf. \cite{intro3,intro4}) has developed rather quickly to be an important one. 

It is very important to get a good understanding of entanglement properties of the quantum states, under effects of accelerations (Lorentz transformations) and magnetic fields (constant and homogeneous), e.g., in studying quantum entangled states of the particles that are produced and detected in Stern-Gerlach experiments \cite{key4}. The issues of quantum entanglement in (constant) external magnetic field were addressed in a previous paper \cite{key1}, through the approach of Wigner rotations of canonical spin. Considering frames of observers related through Lorentz transformations, it was shown that the entanglement is frame \textit{independent} but the violation of Bell's inequality is frame \textit{dependent}. Similar features and correlations (or degrees of entanglement) were noted for spins undergoing precession in a magnetic field in that study. However, the study of quantum entanglement in  magnetic fields was limited to 2-particle states of total spin zero. Here we study entangled 3-particle states in constant external magnetic fields and display the emergence of remarkable \textit{periodic} correlations and density matrices. Note that consequent features of the 2-particle sub-systems can also be extracted from them. To our knowledge, such a study of periodicities in correlations and density matrices is being presented and analysed for the first time. We then introduce an electric field orthogonal to the magnetic field, and obtain the corresponding Wigner rotations of the spin states. In this paper, we also propose a new scheme of systematic classification of the entangled states corresponding to well-known ones in the study of 3-particle entanglements (viz. Greenberger-Horne-Zeilinger $\left| GHZ \right \rangle$, Werner $\left| W \right \rangle$, flipped Werner $\left| \widetilde{W}  \right \rangle$ and their variant states), using a particular coupling of three angular momenta.

The plan of the paper is as follows: We introduce the notations and formalism in Section II. In Sections III and IV, the emerging patterns of periodicity in correlations and density matrices, respectively, are studied explicitly for different entangled states, and particularly for certain initial configuration of the velocities.  In Section V, we study the case with an electric field orthogonal to the magnetic field, linking to the preceding case via Lorentz transformation, and obtain the corresponding Wigner rotations of the spin states. In Section VI, we demonstrate the proposed scheme of systematic classification of the entangled states corresponding to well-known ones in the study of 3-particle entanglements, using the particular coupling of three angular momenta. Finally, we make some  concluding remarks and give an outlook of future directions of work in Section VII.

\section{Spin-1/2 particles in constant magnetic field: Formalism}

In this Section, we describe the formalism and notations:
\begin{enumerate}
\item The unitary transformation matrices acting on the spin states of three particles of spin-1/2 and positive rest mass will be constructed for arbitrary initial velocities $(\overrightarrow{v_{1}},\overrightarrow{v_{2}},\overrightarrow{v_{3}})$ of the particles $(1,2,3)$ respectively and a constant magnetic field $\overrightarrow{B}$, and the basic precession equations (see Ref. \cite{key1}, and review articles \cite{key2,key3}) will be used
and generalized.

\item A particularly simple configuration of the initial velocities is then selected for detailed study (later in Section III). This would permit us to display the contents of the generalized equations, for a few selected cases of particular interest without obscuring the basic features due to a profusion of parameters.

This will be achieved by restricting the initial velocities to a plane orthogonal to the magnetic field $\overrightarrow{B}$  and assuming them to be of equal magnitude, namely by imposing 
\begin{equation}
\overrightarrow{B}.\overrightarrow{v_{i}}=0, \qquad \left| \overrightarrow{v_{i}} \right| = v \qquad (i=1,2,3).
\label{eq1}
\end{equation}

Since $\overrightarrow{B}.\overrightarrow{v}$ and $\left| \overrightarrow{v} \right|$ are constants of motion, the conditions \ref{eq1} will hold for all time $t$.

Eight initial entangled 3-particle states would be selected for detailed study in such a context. They are encoded using the following notations (for our three spin-1/2 particles)

\begin{equation}
 \left| ijk \right \rangle \equiv \left| i \right \rangle \otimes \left| j \right \rangle \otimes \left| k \right \rangle ,
\label{eq2}
\end{equation}
where 

\begin{equation}
 \left| i \right \rangle \subset \LP \left| + \right \rangle, \left| - \right \rangle \RP \equiv \LP \left| 1 \right \rangle, \left| \overline{1} \right \rangle \RP ,
\label{eq3}
\end{equation}
and similarly for $\LP \left| j \right \rangle, \left| k \right \rangle \RP $.

In such notations the states at time $t=0$ are

\begin{equation}
 \frac{1}{\sqrt{2}} \LP \left| 111 \right \rangle + \epsilon \left| \overline{1} \overline{1} \overline{1} \right \rangle \RP ,
\label{eq4}
\end{equation}

\begin{equation}
 \frac{1}{\sqrt{3}} \LP \left| 11\overline{1} \right \rangle + \exp({-i \phi}) \left| 1\overline{1} 1 \right \rangle  + \exp({i \phi}) \left| \overline{1} 11 \right \rangle \RP ,
\label{eq5}
\end{equation}
and
\begin{equation}
 \frac{1}{\sqrt{3}} \LP \left| \overline{1}\overline{1}1 \right \rangle +  \exp({i \phi}) \left| \overline{1} 1  \overline{1} \right \rangle  + \exp({-i \phi}) \left| 1 \overline{1} \overline{1} \right \rangle \RP,
\label{eq6}
\end{equation}
where $\LP  \epsilon = \pm \RP $ and $\LP \phi = 0,\pm \frac{2 \pi}{3}\RP$.

Not only do these states have direct correspondence to the $\left| GHZ \right \rangle, \left| W \right \rangle, \left| \widetilde{W}  \right \rangle$ states familiar in the study of 3-particle entanglements  \cite{key4,key5} (Ref.  \cite{key5} cites original sources), but they have also been constructed long ago \cite{key6,key7}  in the study of coupling of three angular momenta involving eigenvalues of the operator

\begin{equation}
Z= \LP \overrightarrow{J_1} \times \overrightarrow{J_2} \RP . \overrightarrow{J_3},
\label{eq7}
\end{equation}
of three angular momenta $\LP \overrightarrow{J_1} , \overrightarrow{J_2} , \overrightarrow{J_3} \RP$, as is briefly explained in Section VI.

\item We will follow the time-evolution of the states \eq{eq4}-\eq{eq6} with special attention to {\it periodic oscillations} (in Section III), and also study the corresponding {\it periodic density matrices}  (in Section IV), displaying many interesting features.

\item We will generalize the background field to include a constant electric field $\overrightarrow{E}$, orthogonal to the magnetic field $\overrightarrow{B}$, such that
%\begin{equation}
 $$\overrightarrow{E}. \overrightarrow{B}=0, \qquad \left| \overrightarrow{E} \right| < \left| \overrightarrow{B} \right|.$$
%\label{eq7}
%\end{equation}
This constraint would permit us to obtain the results via an appropriate Lorentz transformation (in Section V), and adapt the results from Ref. \cite{key1} (citing original sources) for the present 3-particle case.

\end{enumerate}

%\section{Three particles of spin-1/2 in constant magnetic field}

First, following the lines of Ref. \cite{key1}, for a \textit{single} particle we denote $(\overrightarrow{B},\overrightarrow{v},\overrightarrow{\Sigma})$ to be respectively the constant, homogeneous magnetic field, the velocity and the polarization. With unit vectors $(\hat{B},\hat{v})$ and $c=1$, we have

\begin{equation}
 \overrightarrow{B}= B. \hat{B} \qquad \overrightarrow{v}= v. \hat{v}, \qquad \gamma = \LP 1-v^{2}\RP^{1/2}.
\label{eq8}
\end{equation}
The anomalous magnetic moment is denoted by

\begin{equation}
 \hat{\alpha}= (g-2)/2.
\label{eq9}
\end{equation}

We define, for mass $m$ and charge $e$,
\begin{eqnarray}
 \overrightarrow{\omega} &=&\frac{eB}{m\gamma}\hat{B} \nonumber \\
 \overrightarrow{\Omega} &=&\frac{\hat{\alpha} eB}{m\gamma}\LP \gamma \hat{B}-(\gamma-1)  (\hat{B}.\hat{v}) \hat{v}\RP.
\label{eq10}
\end{eqnarray}

The equations for  $\hat{v}$ and  $\overrightarrow{\Sigma}$ are
\begin{eqnarray}
 \frac{\rm{d}\overrightarrow{v}}{\rm{dt}}&=& -\overrightarrow{\omega} \times \overrightarrow{v} \nonumber \\
 \frac{\rm{d}\overrightarrow{\Sigma}}{\rm{dt}} &=&  -(\overrightarrow{\omega}+ \overrightarrow{\Omega})\times \overrightarrow{\Sigma}.
\label{eq11}
\end{eqnarray}

The constants of motion are $(v,\hat{B}.\hat{v})$ and the moduli

\begin{eqnarray}
 \omega &=& \LP \frac{eB}{m\gamma} \RP \nonumber \\
 \Omega &=& \LP \frac{\hat{\alpha} eB}{m\gamma} \RP \LP \gamma^{2} -(\gamma^{2}-1)  (\hat{B}.\hat{v})^{2} \RP^{1/2}
\label{eq12}
\end{eqnarray}
such that $\overrightarrow{\omega}= \omega \hat{\omega}, \overrightarrow{\Omega}= \Omega \hat{\Omega}$.

Introducing an intermediate set of rotating axes, one can finally obtain the time-dependent unitary transformation matrix acting on spin states  $(\left| \frac{1}{2} \right \rangle, \left| -\frac{1}{2} \right \rangle )$ of a particle moving with velocity $\overrightarrow{v}$ as

\begin{equation}
 M= \exp \LP -i \frac{\Omega t}{2} (\hat{\Omega}.\overrightarrow{\sigma})\RP \exp \LP -i \frac{\omega t}{2} (\hat{B}.\overrightarrow{\sigma})\RP
\label{eq13}
\end{equation}
where $\overrightarrow{\sigma}$ denote the Pauli matrices and the components of  $\overrightarrow{\Omega}$ are 
(assuming $(\hat{B}.\hat{v}) \neq \pm 1$)

\begin{eqnarray}
 \Omega_1 &=& (\hat{B}.\overrightarrow{\Omega}) = \frac{\alpha eB}{m\gamma} \LP \gamma -(\gamma-1)  (\hat{B}.\hat{v})^{2} \RP \nonumber \\
 \Omega_2 &=& \frac{(\hat{B} \times \hat{v}). \overrightarrow{\Omega}}{\sqrt{1-  (\hat{B}.\hat{v})^{2}}} = 0 \nonumber \\
 \Omega_3 &=& \frac{ (\hat{B} \times (\hat{B} \times \hat{v})).\overrightarrow{\Omega}}{\sqrt{1-  (\hat{B}.\hat{v})^{2}}} \nonumber \\
 ~ &=&  \frac{\alpha eB}{m\gamma}(\gamma-1)  (\hat{B}.\hat{v})\sqrt{1-  (\hat{B}.\hat{v})^{2}} .
\label{eq14}
\end{eqnarray}
%For $(\hat{B}.\hat{v}) = \pm 1$, there is no precession.
Below we leave aside the particularly simple case arising for $(\hat B . \hat v ) =\pm 1$.

We follow the same prescription as in Ref. \cite{key1}, and denote
\begin{eqnarray}
 \hat{B}.\overrightarrow{\sigma}  &=& (b_1 \sigma_1+  b_2 \sigma_2+ b_3 \sigma_3) \nonumber \\
 \hat{\Omega}.\overrightarrow{\sigma}  &=& (l_1 \sigma_1+  l_3 \sigma_3) ,
\label{eq15}
\end{eqnarray}
so that
\begin{eqnarray}
 b_1^{2}+  b_2^{2} +  b_3^{2} &=& 1 \nonumber \\
 l_1^{2}+  l_3^{2} &=& 1 , \qquad  l_2=0.
\label{eq15a}
\end{eqnarray}
We also introduce the notations

\begin{eqnarray}
 (c, s) & \equiv & \LP \cos (\frac{\omega t}{2}),  \sin (\frac{\omega t}{2})\RP \nonumber \\
  (c', s') & \equiv & \LP \cos (\frac{\Omega t}{2}),  \sin (\frac{\Omega t}{2})\RP ,
\label{eq16}
\end{eqnarray}
so that we can now write 
%\[ M = \left| \begin{array}{ccc}
%\lambda - a & -b & -c \\
%-d & \lambda - e & -f \\
%-g & -h & \lambda - i \end{array} \right|.\] 

%\begin{eqnarray}
% M = \left| \begin{array}{cc}
% (c'-il_3s') & -il_1s' \\
% -il_1s' & (c'+il_3s')
%\end{array} \right|\left| \begin{array}{cc}
% (c-ib_3s) & -i(b_1-ib_2)s \\
% -i(b_1+ib_2)s & (c+ib_3s)
%\end{array} \right| ,
%\label{eq17}
%\end{eqnarray}
%or
\begin{eqnarray}
 M \equiv \left| \begin{array}{cc}
 \alpha & -i\beta \\
 -i\beta^{*} &  \alpha^{*}
\end{array} \right|,
\label{eq18}
\end{eqnarray}
where we define
\begin{eqnarray}
 \alpha &=& (c'-il_3s')(c-ib_3s)-l_1s'(b_1+ib_2)s \nonumber \\
  \beta  &=& (c'-il_3s')(b_1-ib_2)s +l_1s'(c+ib_3s) ,
\label{eq19}
\end{eqnarray}
which will be the \textit{precession parameters}.

Using the notations defined above, we obtain the unitarity constraint
\begin{equation}
  M^{\dagger}  M \equiv (\alpha \alpha^{*}+ \beta \beta^{*} ) \left| \begin{array}{cc}
 1 & 0 \\
 0 & 1
\end{array} \right|
= \left| \begin{array}{cc}
 1 & 0 \\
 0 & 1
\end{array} \right|.
\label{eq20}
\end{equation}
The action of $M$ on the spin states 
\begin{equation}
 \LP \left| + \right \rangle, \left| - \right \rangle \RP \equiv \LP \left| 1 \right \rangle, \left| \overline{1} \right \rangle \RP = \LP \left| \begin{array}{c}
 1  \\
 0 
\end{array} \right \rangle, \left|  \begin{array}{c}
 0  \\
 1 
\end{array} \right \rangle \RP,
\label{eq21}
\end{equation}
is given by
\begin{equation}
 M \LP \left| 1 \right \rangle, \left| \overline{1} \right \rangle \RP = \LP (\alpha \left| 1 \right \rangle - i \beta^{*} \left| \overline{1} \right \rangle ), (-i \beta  \left| 1 \right \rangle + \alpha^{*}  \left| \overline{1} \right \rangle )\RP .
\label{eq22}
\end{equation}

So far we have been considering a single particle with velocity $\overrightarrow{v}$. For three particles with velocities $(\overrightarrow{v_{1}},\overrightarrow{v_{2}},\overrightarrow{v_{3}})$ we denote the corresponding precession parameters as 
$$(\alpha_1, \beta_1), (\alpha_2, \beta_2), (\alpha_3, \beta_3)$$
generalizing appropriately
$(\alpha, \beta)$ of \eq{eq19}.

Thus, for example, an inital state at $t=0$, 

\begin{equation}
  \left| A \right \rangle \equiv \frac{1}{\sqrt{2}} \LP \left| 111 \right \rangle + \left| \overline{1} \overline{1} \overline{1} \right \rangle \RP  ,
\label{eq23}
\end{equation}
will evolve as
\begin{widetext}
\begin{eqnarray}
 \LP M_{(1)} \otimes  M_{(2)}  \otimes  M_{(3)} \RP  \left| A \right \rangle =  \frac{1}{\sqrt{2}} (\alpha_1 \left| 1 \right \rangle - i \beta_1^{*} \left| \overline{1} \right \rangle ) (\alpha_2 \left| 1 \right \rangle - i \beta_2^{*} \left| \overline{1} \right \rangle )(\alpha_3 \left| 1 \right \rangle - i \beta_3^{*} \left| \overline{1} \right \rangle ) \nonumber \\+ \frac{1}{\sqrt{2}}  (-i \beta_1  \left| 1 \right \rangle + \alpha_1^{*}  \left| \overline{1} \right \rangle )(-i \beta_2  \left| 1 \right \rangle + \alpha_2^{*}  \left| \overline{1} \right \rangle )(-i \beta_3  \left| 1 \right \rangle + \alpha_3^{*}  \left| \overline{1} \right \rangle ),
\label{eq24}
\end{eqnarray}
or
\begin{eqnarray}
M_{(1)} \otimes  M_{(2)}  \otimes  M_{(3)}  \left| A \right \rangle 
&=& d_{111}\left| 111 \right \rangle + d_{1 \overline{1} \overline{1}}\left| 1 \overline{1} \overline{1} \right \rangle +d_{\overline{1} 1 \overline{1}}\left| \overline{1} 1 \overline{1} \right \rangle +d_{\overline{1} \overline{1} 1}\left| \overline{1} \overline{1} 1 \right \rangle \nonumber \\
&+&d_{\overline{1} \overline{1} \overline{1}} \left| \overline{1} \overline{1} \overline{1}\right \rangle +d_{\overline{1} 11}\left| \overline{1} 11 \right \rangle +d_{1 \overline{1} 1}\left| 1 \overline{1} 1 \right \rangle +d_{11\overline{1}}\left| 11\overline{1} \right \rangle,
\label{eq25}
\end{eqnarray}
\end{widetext}
where
\begin{eqnarray}
d_{111} = \frac{1}{\sqrt{2}} (\alpha_1 \alpha_2 \alpha_3 +i \beta_1 \beta_2 \beta_3) \nonumber \\
d_{1 \overline{1} \overline{1}}  = -\frac{1}{\sqrt{2}} (\alpha_1\beta_2^{*} \beta_3^{*} +i\beta_1 \alpha_2^{*} \alpha_3^{*}) \nonumber \\
d_{\overline{1} 1 \overline{1}}  = -\frac{1}{\sqrt{2}} ( \beta_1^{*}\alpha_2 \beta_3^{*}+i \alpha_1^{*} \beta_2 \alpha_3^{*}) \nonumber \\
d_{\overline{1} \overline{1} 1} = -\frac{1}{\sqrt{2}} ( \beta_1 \beta_2 \alpha_3+i\alpha_1 \alpha_2 \beta_3) \nonumber \\
d_{\overline{1} \overline{1} \overline{1}}  = \frac{1}{\sqrt{2}} (\alpha_1^{*} \alpha_2^{*} \alpha_3^{*} +i \beta_1^{*} \beta_2^{*} \beta_3^{*}) \nonumber \\
d_{\overline{1} 11}  = -\frac{1}{\sqrt{2}} (i\beta_1^{*} \alpha_2 \alpha_3+\alpha_1^{*} \beta_2 \beta_3) \nonumber \\
d_{1 \overline{1} 1}  = -\frac{1}{\sqrt{2}} (i \alpha_1 \beta_2^{*}\alpha_3 + \beta_1 \alpha_2^{*} \beta_3) \nonumber \\
d_{11\overline{1}} = -\frac{1}{\sqrt{2}} (i\alpha_1 \alpha_2  \beta_3^{*}+ \beta_1 \beta_2 \alpha_3^{*}).
\label{eq26}
\end{eqnarray}

The coefficients above now involve \textit{three distinct periodic time dependences} through
\begin{eqnarray}
 (c, s)_{i} & \equiv & \LP \cos (\frac{\omega_{i} t}{2}),  \sin (\frac{\omega_{i} t}{2})\RP \nonumber \\
  (c', s')_{i} & \equiv & \LP \cos (\frac{\Omega_{i} t}{2}),  \sin (\frac{\Omega_{i} t}{2})\RP,
\label{eq27}
\end{eqnarray}
where $(i=1,2,3)$, determined by the initial velocities $(\overrightarrow{v_{1}},\overrightarrow{v_{2}},\overrightarrow{v_{3}})$  and their orientations with respect to $\overrightarrow{B}$. Note that we ignore the mutual interactions of the particles assumed to be weak enough as compared to that with a strong magnetic field $\overrightarrow{B}$.

The probability associated to the state $\left| ijk \right \rangle $ is defined to be
\begin{equation}
  P_{ijk}=d_{ijk}^*d_{ijk} .
\label{eq28}
\end{equation}
We note that in $\left| ijk \right \rangle $,  the sum of the probablities of $\left| i \right \rangle $ being either $\left| 1 \right \rangle $ or $\left| \overline{1} \right \rangle $ must be $1$. Using 
the relation 
 \begin{equation}
  (\alpha_i \alpha_i^{*}+ \beta_i \beta_i^{*} )=1 \qquad (i=1,2,3)
\label{eq29}
\end{equation}
systematically, one may check that, for example,

\begin{eqnarray}
  P_{11} &=& P_{111}+P_{11\overline{1}} \nonumber \\
   &=& \frac{1}{2}(\alpha_1 \alpha_1^{*}\alpha_2 \alpha_2^{*}+ \beta_1 \beta_1^{*}\beta_2 \beta_2^{*} ) \\ 
  P_{1\overline{1}} &=& \frac{1}{2}(\alpha_1 \alpha_1^{*}\beta_2 \beta_2^{*}+ \beta_1 \beta_1^{*}\alpha_2 \alpha_2^{*} ) \\
  P_{1} &=& P_{11}+P_{1\overline{1}} \nonumber \\
   &=& \frac{1}{2}(\alpha_1 \alpha_1^{*}+ \beta_1 \beta_1^{*})(\alpha_2 \alpha_2^{*}+\beta_2 \beta_2^{*} ) \nonumber \\
    &=& \frac{1}{2}.
\label{eq30}
\end{eqnarray}
Similarly,
\begin{equation}
   P_{\overline{1}} = P_{\overline{1}1}+P_{\overline{1}\overline{1}} = \frac{1}{2}.
\label{eq31}
\end{equation}
Hence
\begin{equation}
  P_{1}+ P_{\overline{1}}=1 .
\label{eq32}
\end{equation}
Analogous results hold for $(j,k)$.

For the general configuration the interplay of the periods (see \eq{eq27})

\begin{equation}
   \LP \frac{4\pi}{\omega_i}, \frac{4\pi}{\Omega_i} \RP \qquad (i=1,2,3)
\label{eq33}
\end{equation}
imply a \textit{rich} structure in the variation with $t$ of the coefficients $d_{ijk}$ in \eq{eq25} and the correlations  \eq{eq28}. Such variations will of course depend on the velocities and masses involved.

\section{Periodic correlations for  3-particle states in a  magnetic field}

In this section we will select, to start with, a simple case and try to follow closely the periodicities associated with the intial entangled 3-particle states given by \eq{eq4}-\eq{eq6}.

The constant magnetic field $\overrightarrow{B}$ is taken to be along the x-axis,
\begin{equation}
   \overrightarrow{B}= (B,0,0).
\label{eq3_1}
\end{equation}
Three equal mass spin-1/2 particles (created, say, by the disintegration of a single one at rest) are assumed to have their intial velocities in the yz-plane and to be of equal magnitude, so that
\begin{equation}
   \overrightarrow{B}.\overrightarrow{v_{(i)}}=0, \qquad \left| \overrightarrow{v_{(i)}}  \right|=v \qquad (i=1,2,3).
\label{eq3_2}
\end{equation}

Since $\overrightarrow{v}^2$ and $\overrightarrow{B}.\overrightarrow{v}$ are constants of motion, the velocities wil stay in the yz-plane and remain of equal magnitude. The three velocities rotate uniformly in the yz-plane.

The precession equations \eq{eq11} now simplify (for each case) to

$$ \frac{\rm{d}\overrightarrow{\Sigma}}{\rm{dt}} =  -(\overrightarrow{\omega}+ \overrightarrow{\Omega})\times \overrightarrow{\Sigma},$$
where
\begin{eqnarray}
\qquad \overrightarrow{\omega} &=&\frac{eB}{m\gamma}\hat{B} \equiv \omega \hat{B} \nonumber \\
 \overrightarrow{\Omega} &=&\frac{\hat{\alpha} eB}{m} \hat{B}=\hat{\alpha} \gamma \omega \hat{B} \equiv \Omega \hat{B}.
\label{eq3_3}
\end{eqnarray}
and hence
\begin{eqnarray}
  \frac{\rm{d}\overrightarrow{\Sigma}}{\rm{dt}} &=&  -(\omega+ \Omega)\hat{B} \times \overrightarrow{\Sigma}.
\label{eq3_4}
\end{eqnarray}

Now in \eq{eq24} the unitary transformation acting on a state  $\left| ijk \right \rangle$ is given by the matrix
$$M \otimes  M  \otimes  M $$
where 
\begin{equation}
 M \equiv \left| \begin{array}{cc}
 \alpha & -i\beta \\
 -i\beta &  \alpha
\end{array} \right|,
\label{eq3_5}
\end{equation}
and
\begin{eqnarray}
 \alpha &=& \cos (\frac{\omega+ \Omega)}{2}t) \nonumber \\
  \beta  &=& \sin (\frac{\omega+ \Omega)}{2}t) .
\label{eq3_6}
\end{eqnarray}

\subsection{Case 1}
We start by studying the time-evolution of the state \eq{eq4}. One obtains at time $t$

\begin{align}
M \otimes  M  \otimes  M  \frac{1}{\sqrt{2}} \LP \left| 111  \right \rangle +\epsilon \left|  \overline{1} \overline{1} \overline{1}  \right \rangle \RP
= f_{111}\left| 111 \right \rangle \nonumber \\
+ f_{1 \overline{1} \overline{1}}\left| 1 \overline{1} \overline{1} \right \rangle +f_{\overline{1} 1 \overline{1}}\left| \overline{1} 1 \overline{1} \right \rangle 
+f_{\overline{1} \overline{1} 1}\left| \overline{1} \overline{1} 1 \right \rangle 
+f_{\overline{1} \overline{1} \overline{1}} \left| \overline{1} \overline{1} \overline{1}\right \rangle \nonumber \\ +f_{\overline{1} 11}\left| \overline{1} 11 \right \rangle 
+f_{1 \overline{1} 1}\left| 1 \overline{1} 1 \right \rangle +f_{11\overline{1}}\left| 11\overline{1} \right \rangle,
\label{eq3_7}
\end{align}

where
\begin{eqnarray}
f_{111} = \frac{1}{\sqrt{2}} (\alpha^3 +i \epsilon \beta^3) = \epsilon f_{\overline{1} \overline{1} \overline{1}}
\label{eq3_8}
\end{eqnarray}
and
\begin{align}
f_{1 \overline{1} \overline{1}}  = f_{\overline{1} 1 \overline{1}}  = f_{\overline{1} \overline{1} 1} = f_{\overline{1} 11}  = f_{1 \overline{1} 1}  = f_{11\overline{1}} \nonumber \\
= -\frac{1}{\sqrt{2}} \alpha  \beta( \alpha + \epsilon i \beta).
\label{eq3_9}
\end{align}

The corresponding probabilities given by $P_{ijk}=f_{ijk}^*f_{ijk}$ are
\begin{align}
P_{111} &=& P_{\overline{1} \overline{1} \overline{1}} = \frac{1}{2} (1-3\alpha^2 \beta^2)  \nonumber \\
~&=& \frac{1}{2} (1- \frac{3}{4} (\sin^2(( \omega+ \Omega)t)))
\label{eq3_10}
\end{align}
\begin{align}
P_{1 \overline{1} \overline{1}}  = P_{\overline{1} 1 \overline{1}}  = P_{\overline{1} \overline{1} 1} = P_{\overline{1} 11}  = P_{1 \overline{1} 1}  = P_{11\overline{1}} \nonumber \\
= \frac{1}{8}(\sin^2(( \omega+ \Omega)t)) ,
\label{eq3_11}
\end{align}
consistent with the constraint of
\begin{equation}
 \sum_{ijk} P_{ijk}=1.
\label{eq3_12}
\end{equation}

\begin{figure}[h]
\includegraphics[angle=270,width=0.45\textwidth]{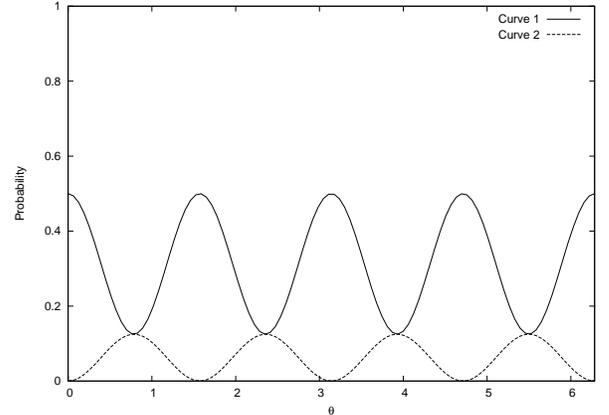}
\caption{Plot of the variations of the probabilities $P_{ijk}$ defined in the text by \eq{eq3_10} (Curve 1) and \eq{eq3_11} (Curve 2) with $\theta \equiv (\frac{\omega+ \Omega)}{2}t)$.}
\end{figure}

Figure 1 shows the variations of the probabilities $P_{ijk}$ defined in \eq{eq3_10} and \eq{eq3_11} with $\theta \equiv (\frac{\omega+ \Omega)}{2}t)$. The periodic variations of the coefficients are easy to follow. Let us 
emphasize some special points:
\begin{enumerate}
\item At $t=0$, one starts with $\alpha=1,  \beta=0$
\begin{equation}
 \left| \psi \right \rangle \equiv \sum_{ijk} f_{ijk}\left| ijk \right \rangle=\frac{1}{\sqrt{2}} \LP \left| 111  \right \rangle +\epsilon \left|  \overline{1} \overline{1} \overline{1}  \right \rangle \RP.
\label{eq3_13}
\end{equation}
\item At $t=\frac{\pi}{2( \omega+ \Omega)}$, ($\alpha= \beta= \frac{1}{\sqrt{2}}$)
\begin{align}
\left| \psi \right \rangle \equiv \frac{\exp{(i \epsilon \pi /4)} }{2\sqrt{2}} \LB \LP \left| 111  \right \rangle +\epsilon \left|  \overline{1} \overline{1} \overline{1}  \right \rangle \RP -  \LP \left| 1 \overline{1} \overline{1} \right \rangle +\left| \overline{1} 1 \overline{1} \right \rangle 
+\left| \overline{1} \overline{1} 1 \right \rangle \RP  - \LP \left| \overline{1} 11 \right \rangle 
+ \left| 1 \overline{1} 1 \right \rangle +\left| 11\overline{1} \right \rangle \RP \RB.
\label{eq3_14}
\end{align}
Note the negative signs before the two triplets, whose consequences will be discussed in Section IV.

Now all the 8 possible states are present with equal probability
\begin{equation}
 P_{ijk}= \frac{1}{8}  \qquad (i,j,k)=1 \quad {\rm or} \quad \overline{1}.
\label{eq3_15}
\end{equation}

\item At $t=\frac{\pi}{( \omega+ \Omega)}$, ($\alpha= 0 , \beta= 1$)
\begin{equation}
 \left| \psi \right \rangle =\frac{\epsilon }{\sqrt{2}} \LP \left| 111  \right \rangle +\epsilon \left|  \overline{1} \overline{1} \overline{1}  \right \rangle \RP.
\label{eq3_16}
\end{equation}
Note that apart from an overall sign ($\epsilon$) one is back at the starting point.
\end{enumerate}

\subsection{Case 2}
We will now study the time-evolution with the next initial state given by \eq{eq5}:
%\begin{widetext}
\begin{align}
 \left| \psi \right \rangle_{(t=0)} &= \left| \psi \right \rangle_{(0)} \nonumber \\
 &= \frac{1}{\sqrt{3}} \LP \left| 11\overline{1} \right \rangle + \exp({-i \phi}) \left| 1\overline{1} 1 \right \rangle  + \exp({i \phi}) \left| \overline{1} 11 \right \rangle \RP , 
\label{eq3_17}
\end{align}
%\end{widetext}
where $ \qquad \LP \phi = 0, \pm \frac{2 \pi}{3}\RP. $
We define
\begin{eqnarray}
 f &=& (1+\exp({-i \phi})+\exp({i \phi})) \nonumber \\
   &=& 3  \qquad \LP \phi = 0\RP \nonumber \\
   &=& 0  \qquad \LP \phi = \pm \frac{2 \pi}{3}\RP .
\label{eq3_18}
\end{eqnarray}
At time $t$, the periodic evolution gives (using \eq{eq3_5})
\begin{eqnarray}
\left| \psi \right \rangle_{(t)} &=& M \otimes  M  \otimes  M \left| \psi \right \rangle_{(0)} \nonumber \\
   &=& c_0 \left| 111  \right \rangle + \overline{c_0} \left|  \overline{1} \overline{1} \overline{1}  \right \rangle  \nonumber \\
   &+& c_1 \left| 11\overline{1} \right \rangle + c_2 \left| 1 \overline{1} 1 \right \rangle + c_3 \left| \overline{1} 11 \right \rangle  \nonumber \\
   &+& \overline{c_1} \left| \overline{1} \overline{1} 1 \right \rangle+ \overline{c_2} \left| \overline{1} 1 \overline{1} \right \rangle + \overline{c_3} \left| 1 \overline{1} \overline{1} \right \rangle.
\label{eq3_19}
\end{eqnarray}
where with
$$\alpha = \cos (\frac{\omega+ \Omega)}{2}t), \beta  = \sin (\frac{\omega+ \Omega)}{2}t)$$
and implementing systematically $\alpha^2 +\beta^2  = 1$, we have
\begin{align}
 \sqrt3 c_0 = -i \alpha^2 \beta f &,& \sqrt3 \overline{c_0} = - \alpha \beta^2 f\nonumber \\
 \sqrt3 c_1 = \alpha(1 -\beta^2f) &,& \sqrt3 \overline{c_1} = i \beta(1- \alpha^2f)\nonumber \\
 \sqrt3 c_2 = \alpha(\exp({-i \phi})-\beta^2f)  &,& \sqrt3 \overline{c_2} = i \beta(\exp({-i \phi})- \alpha^2f) \nonumber \\
 \sqrt3 c_3 = \alpha(\exp({i \phi})-\beta^2f)  &,& \sqrt3 \overline{c_3} = i \beta(\exp({i \phi})- \alpha^2f)   .
\label{eq3_20}
\end{align}
The correlations have periodicities determined by ($\alpha, \beta$) above and are
\begin{eqnarray}
P_{111} = \frac{1}{3} \alpha^4\beta^2f^2 ,  P_{\overline{1} \overline{1} \overline{1}} = \frac{1}{3} \alpha^2\beta^4f^2  &,& \nonumber \\
P_{1 \overline{1} \overline{1}}  = \frac{1}{3} \beta^2(1-\alpha^2f\exp({-i \phi}))(1-\alpha^2f\exp({i \phi})) &,& \nonumber \\
P_{\overline{1} 11}  = \frac{1}{3} \alpha^2(1-\beta^2f\exp({-i \phi}))(1-\beta^2f\exp({i \phi})) &,& \nonumber \\
P_{1 \overline{1} 1} = \frac{1}{3} \alpha^2(1-\beta^2f\exp({-i \phi}))(1-\beta^2f\exp({i \phi}))  &,&
\nonumber \\
 P_{\overline{1} 1 \overline{1}} = \frac{1}{3} \beta^2(1-\alpha^2f\exp({-i \phi}))(1-\alpha^2f\exp({i \phi})) &,& \nonumber \\
P_{\overline{1} \overline{1} 1} = \frac{1}{3} \beta^2(1-\alpha^2f)^2,
P_{11\overline{1}} = \frac{1}{3} \alpha^2(1-\beta^2f)^2 .
\label{eq3_21}
\end{eqnarray}
After this derivation, one notes that for:
\begin{enumerate}
\item ($\phi= 0 , f= 3$)
\begin{eqnarray}
P_{111} = 3 \alpha^4\beta^2 &,& P_{\overline{1} \overline{1} \overline{1}} = 3 \alpha^2\beta^4  \nonumber \\
P_{\overline{1} \overline{1} 1} =  P_{1 \overline{1} \overline{1}}  = P_{\overline{1} 1 \overline{1}}= \frac{1}{3} \beta^2(1-3\alpha^2)^2 \nonumber \\
P_{\overline{1} 11} =P_{1 \overline{1} 1} = P_{11\overline{1}} = \frac{1}{3} \alpha^2(1-3\beta^2)^2 .
\label{eq3_22}
\end{eqnarray}
consistent with the constraint \eq{eq3_12} of 
$\sum_{ijk} P_{ijk}=1.$
%\begin{equation}
%$\sum_{ijk} P_{ijk}=1.$
%\label{eq3_12}
%\end{equation}

\begin{figure}[h]
\includegraphics[angle=270,width=0.45\textwidth]{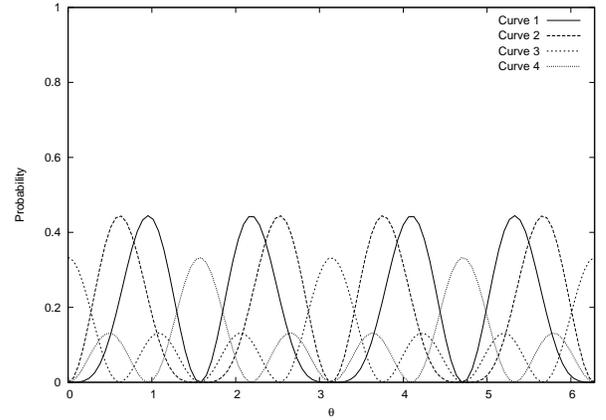}
\caption{Plot of the variations of the probabilities $P_{ijk}$ defined in the text by \eq{eq3_22} with $\theta \equiv (\frac{\omega+ \Omega)}{2}t)$, where $P_{111}$ is Curve 1, $P_{\overline{1} \overline{1} \overline{1}}$ is Curve 2, $P_{\overline{1} \overline{1} 1} =  P_{1 \overline{1} \overline{1}}  = P_{\overline{1} 1 \overline{1}}$ are Curve 3 and $P_{\overline{1} 11} =P_{1 \overline{1} 1} = P_{11\overline{1}}$ are Curve 4.}
\end{figure}

Figure 2 shows the variations of the probabilities $P_{ijk}$ defined in \eq{eq3_22} with $\theta \equiv (\frac{\omega+ \Omega)}{2}t)$.

\item ($\phi= \pm \frac{2\pi}{3} , f= 0$)
\begin{eqnarray}
P_{111} =  P_{\overline{1} \overline{1} \overline{1}} = 0  \nonumber \\
P_{\overline{1} \overline{1} 1} =  P_{1 \overline{1} \overline{1}}  = P_{\overline{1} 1 \overline{1}}= \frac{1}{3} \beta^2 \nonumber \\
P_{\overline{1} 11} =P_{1 \overline{1} 1} = P_{11\overline{1}} = \frac{1}{3} \alpha^2 .
\label{eq3_23}
\end{eqnarray}
consistent with the constraint \eq{eq3_12} of 
$\sum_{ijk} P_{ijk}=1.$
%\begin{equation}
% \sum_{ijk} P_{ijk}=1.\nonumber \\
%%\label{eq3_12}
%\end{equation}

\end{enumerate}
Case (i) exhibits a richer pattern of periodicity which we analyse now, pinpoiting some special values of $t$.
 At $t=0$, one starts with $\alpha=1,  \beta=0$:
\begin{equation}
 \left| \psi \right \rangle_{(0)} =\frac{1}{\sqrt{3}} \LP \left| 11 \overline{1}  \right \rangle + \left| 1 \overline{1} 1  \right \rangle  + \left| \overline{1}11  \right \rangle \RP.
\label{eq3_24}
\end{equation}

At $t=\frac{\pi}{( \omega+ \Omega)}$, ($\alpha= 0 , \beta= 1$):
\begin{equation}
 \left| \psi \right \rangle =\frac{i }{\sqrt{3}} \LP \left| \overline{1}\overline{1}1  \right \rangle + \left|  \overline{1}1 \overline{1} \right \rangle + \left| 1 \overline{1} \overline{1} \right \rangle\RP.
\label{eq3_25}
\end{equation}
we get the flipped over complementary triplet.

At $t=\frac{\pi}{2( \omega+ \Omega)}$, ($\alpha= \beta= \frac{1}{\sqrt{2}}$):
\begin{eqnarray}
\left| \psi \right \rangle = \frac{-i}{2\sqrt{6}} \LP 3 \left| 111  \right \rangle + \left|  \overline{1} \overline{1} 1  \right \rangle + \left| \overline{1} 1 \overline{1} \right \rangle +\left| 1 \overline{1} \overline{1} \right \rangle \RP \nonumber \\
+ \frac{1}{2\sqrt{6}}  \LP 3 \left| \overline{1} \overline{1} \overline{1} \right \rangle - \left| 11\overline{1} \right \rangle -\left| 1 \overline{1} 1 \right \rangle -\left| \overline{1}11 \right \rangle \RP.
\label{eq3_26}
\end{eqnarray}
%consistently with the \eq{eq3_12}
%\begin{equation}
% \sum_{ijk} P_{ijk}=1.\nonumber \\
%\label{eq3_12}
%\end{equation}
Now all the possible $\left| ijk \right \rangle$  (8 in number) appear, grouped as above according as the multiplicity of the index $\left| 1 \right \rangle$ is odd (3 or 1) or even (0 or 2).

Other interesting points are provided by the extrema of ($P_{111} ,  P_{\overline{1} \overline{1} \overline{1}}$).
For $\alpha^2=2/3,  \beta^2=1/3$,

\begin{eqnarray}
P_{111} = \frac{4}{9},  P_{\overline{1} \overline{1} \overline{1}} = \frac{2}{9}  \nonumber \\
P_{\overline{1} \overline{1} 1} =  P_{1 \overline{1} \overline{1}}  = P_{\overline{1} 1 \overline{1}}= \frac{1}{9} \nonumber \\
P_{\overline{1} 11} =P_{1 \overline{1} 1} = P_{11\overline{1}} = 0.
\label{eq3_27}
\end{eqnarray}
Considering the positive roots for example, $\alpha=\sqrt{2/3},  \beta=1/\sqrt{3}$,
\begin{equation}
\left| \psi \right \rangle = \frac{-i}{3} \LP 2 \left| 111  \right \rangle + \left|  \overline{1} \overline{1} 1  \right \rangle + \left| \overline{1} 1 \overline{1} \right \rangle +\left| 1 \overline{1} \overline{1} \right \rangle \RP - \frac{\sqrt{2}}{3}  \left| \overline{1} \overline{1} \overline{1} \right \rangle .
\label{eq3_28}
\end{equation}

For $\alpha=1/\sqrt{3},  \beta=\sqrt{2/3}$,
\begin{eqnarray}
P_{111} = \frac{2}{9},  P_{\overline{1} \overline{1} \overline{1}} = \frac{4}{9}  \nonumber \\
P_{\overline{1} \overline{1} 1} =  P_{1 \overline{1} \overline{1}}  = P_{\overline{1} 1 \overline{1}}= 0 \nonumber \\
P_{\overline{1} 11} =P_{1 \overline{1} 1} = P_{11\overline{1}} = \frac{1}{9},
\label{eq3_29}
\end{eqnarray}
and
\begin{equation}
\left| \psi \right \rangle = \frac{-i\sqrt{2}}{3}  \left| 111  \right \rangle - \frac{1}{3} \LP 2 \left| \overline{1} \overline{1} \overline{1} \right \rangle + \left| 11\overline{1} \right \rangle + \left| 1 \overline{1} 1 \right \rangle + \left| \overline{1}11 \right \rangle\RP.
\label{eq3_30}
\end{equation}
Apart from the factor $i$ changing place, the passage from \eq{eq3_28} to \eq{eq3_30} corresponds to

\begin{equation}
(\left| 1 \right \rangle, \left| \overline{1} \right \rangle) \rightarrow (\left| \overline{1} \right \rangle, \left| 1 \right \rangle).
\label{eq3_31}
\end{equation}

The coefficients for the cases above will be further discussed in Section IV.

The time-evolution of the initial state
\begin{equation}
 \left| \psi \right \rangle_{(0)} = \frac{1}{\sqrt{3}} \LP \left| \overline{1}\overline{1}1 \right \rangle + \exp({i \phi}) \left| \overline{1} 1 \overline{1}\right \rangle  + \exp({-i \phi}) \left| 1\overline{1}\overline{1} \right \rangle \RP 
\label{eq3_32}
\end{equation}
can be studied in closely analogous fashion, but we will not repeat the steps again.

The periods considered above, typically generated by ($\alpha,  \beta$), of \eq{eq3_6} where ($\omega,  \Omega$) are given by  \eq{eq12} can be long but diminish as $B$ and the velocities increase in magnitude.

\section{Periodic density matrices}

We have seen how, starting with a restricted initial state, spin precessions in a magnetic field lead to periodic appearances and disappearances of all the possible eight states $\left| ijk \right \rangle$ of three spin-1/2 particles . Since such periodicities are induced through ``local'' unitary transformations, acting separately on each state $\left| \pm \right \rangle$, the sum of the varying correlations remains unity. Moreover, \textit{the basic invariant measures of entanglement (3-tangle, 2-tangles) must also be conserved}. The corresponding \textit{constrained} periodic variations of the elements of the density matrices is briefly studied below for special cases.

For 
\begin{eqnarray}
\left| \psi \right \rangle = \LP f_0\left| 11  \right \rangle+  f_1\left| 1\overline{1} \right \rangle+ f_2\left|\overline{1}1  \right \rangle+ f_3\left| \overline{1}\overline{1} \right \rangle\RP \left| 1 \right \rangle \nonumber \\
+ \LP g_0\left| \overline{1}\overline{1} \right \rangle+ g_1\left| \overline{1}1 \right \rangle+ g_2\left| 1\overline{1} \right \rangle+ g_3\left| 11 \right \rangle \RP \left| \overline{1} \right \rangle
\label{eq4_1}
\end{eqnarray}
tracing out the index 3, the density matrix for the $(12)$ subsystem is
\begin{equation}
\rho_{12}= \left| \begin{array}{cccc}
 a_{00} & a_{01} & a_{02} & a_{03}\\
 a_{01}^{*} & a_{11} & a_{12} & a_{13}\\
 a_{02}^{*} & a_{12}^{*} & a_{22} & a_{23}\\
 a_{03}^{*} & a_{13}^{*} & a_{23}^{*} & a_{33}
\end{array} \right|
\label{eq4_2}
\end{equation}
where
\begin{eqnarray}
 a_{00}=f_0f_0^{*}+g_3g_3^{*},\qquad  a_{01}=f_0f_1^{*}+g_3g_2^{*} \nonumber\\
 a_{02}=f_0f_2^{*}+g_3g_1^{*},\qquad  a_{03}=f_0f_3^{*}+g_3g_0^{*} \nonumber\\
 a_{11}=f_1f_1^{*}+g_2g_2^{*},\qquad  a_{12}=f_1f_2^{*}+g_2g_1^{*} \nonumber\\
 a_{13}=f_1f_3^{*}+g_2g_0^{*},\qquad  a_{22}=f_2f_2^{*}+g_1g_1^{*} \nonumber\\
 a_{23}=f_2f_3^{*}+g_1g_0^{*},\qquad  a_{33}=f_3f_3^{*}+g_0g_0^{*}.
\label{eq4_3}
\end{eqnarray}
Similarly, one can obtain $\rho_{23},\rho_{31}$.

The 3-tangle (invariant under permutations of particles $1,2,3$) is obtained as follows \cite{key8}:

We define 
\begin{equation}
\widetilde{\rho_{12}}= \left| \begin{array}{cc}
 0 & -i\\
 i & 0
\end{array} \right| \otimes
\left| \begin{array}{cc}
 0 & -i\\
 i & 0
\end{array} \right| \rho_{12}^{*}
\left| \begin{array}{cc}
 0 & -i\\
 i & 0
\end{array} \right| \otimes
\left| \begin{array}{cc}
 0 & -i\\
 i & 0
\end{array} \right| .
\label{eq4_4}
\end{equation}
For the subsystems considered ($\rho_{12}\widetilde{\rho_{12}}$) has at most two non-zero eigenvalues.

Let ($\lambda_{2},\lambda_{2}$) be the square roots (positive) of these two. Then the 3-tangle is given by 
\begin{equation}
\tau_{123}=4\lambda_{1}\lambda_{2}.
\label{eq4_5}
\end{equation}
This can also be directly expressed in terms of the coefficients of \eq{eq4_1} above: 
\begin{equation}
\tau_{123}=4\left|d_1-2d_2+4d_3 \right|,
\label{eq4_6}
\end{equation}
where (in our notations of \eq{eq4_1})
\begin{eqnarray}
 d_1 &=& (f_0g_0)^{2}+(f_1g_1)^{2}+(f_2g_2)^{2}+(f_3g_3)^{2} \nonumber\\
 d_2 &=& f_0g_0 (f_1g_1+f_2g_2+f_3g_3)\nonumber\\ 
&+&(f_1g_1f_2g_2+f_2g_2f_3g_3+f_3g_3f_1g_1)\nonumber\\
 d_3 &=& f_0f_3g_1g_2+f_1f_2g_0g_3.
\label{eq4_7}
\end{eqnarray}
The conversion of \eq{eq4_5} to \eq{eq4_6} permits us to relate directly certain central features of the periodicities studied in Section III to constraints imposed by the invariance of \eq{eq4_5}. In particular, let us now evaluate the crucial role of the negative signs signalled below \eq{eq3_14}.

For the initial state \eq{eq3_13}, at $t=0$
\begin{eqnarray}
 f_0= \epsilon g_0 = \frac{1}{\sqrt2} \nonumber\\
 f_1=f_2=f_3=g_1=g_2=g_3=0
\label{eq4_8}
\end{eqnarray}
and thus 
\begin{equation}
d_1 = \frac{1}{4},  d_2=d_3=0.
\label{eq4_9}
\end{equation}
Hence from \eq{eq4_6} 
\begin{equation}
\tau_{123}=1
\label{eq4_10}
\end{equation}
a well-known result for GHZ states \eq{eq3_13}.

At $t=\frac{\pi}{2(\omega+\Omega)}$, from \eq{eq3_14} and \eq{eq4_7},
\begin{equation}
(d_1,d_2,d_3) = \frac{\exp{i\pi/4}}{{(2\sqrt2})^{4}}(4,6,-2).
\label{eq4_11}
\end{equation}
Hence again (as expected),
\begin{equation}
\tau_{123}= \frac{4}{64}\left| 4-2.6-4.2 \right|=1.
\label{eq4_12}
\end{equation}
If all the terms have the same sign, as in 

\begin{eqnarray}
\left| \psi \right \rangle' &=& \frac{1}{2\sqrt2} \LP \left| 111  \right \rangle+  \left| \overline{1}\overline{1}\overline{1} \right \rangle+ \left|1\overline{1}\overline{1}  \right \rangle+ \left| \overline{1}1\overline{1} \right \rangle \RP \nonumber \\
&+& \frac{1}{2\sqrt2} \LP \left| \overline{1}\overline{1}1 \right \rangle 
+ \left| \overline{1}11 \right \rangle + \left| 1\overline{1}1 \right \rangle+ \left| 11\overline{1} \right \rangle \RP
\label{eq4_13}
\end{eqnarray}
then $d_3$ changes sign and 
\begin{equation}
\tau_{123}= \frac{4}{64}\left| 4-2.6+4.2 \right|=0.
\label{eq4_14}
\end{equation}
Thus the two negative signs in \eq{eq3_14} alters $\tau_{123}$ from a minimum (0) to  maximum (1).

For the initial state \eq{eq3_17}, using \eq{eq3_18}-\eq{eq3_20} the f's in \eq{eq4_1} have cubic periodic terms given by ($\alpha\beta^{2},\alpha^{2}\beta$), where
\begin{equation}
\alpha=\cos\LP \frac{\omega+\Omega)}{2}t\RP ,\qquad \beta= \sin \LP \frac{\omega+\Omega)}{2}t\RP.
\label{eq4_15}
\end{equation}

Hence $P_{12}$ in \eq{eq4_2} and \eq{eq4_3} has sixth order periodic terms ($\alpha^{2}\beta^{4},\alpha^{4}\beta^{2}, \alpha^{3}\beta^{3}$). Such periodicities indeed turn out to be compatible with \cite{key9}:
\begin{equation}
\tau_{123}= 0, \tau_{12}=\tau_{13}=\tau_{23}=\frac{4}{9}.
\label{eq4_16}
\end{equation}

One realizes now more fully how elaborate and subtle patterns of periodicities are compatible with constraints of \textit{local} unitary transformations involved in spin-precessions.

\section{Constant orthogonal electric and magnetic fields}

We briefly indicate below how the periodicities studied so far, for a magnetic field alone are affected by the presence of an electric field $\overrightarrow{E}$ satisfying
 
\begin{equation}
 \overrightarrow{E}. \overrightarrow{B}=0, \qquad \left| \overrightarrow{E} \right| < \left| \overrightarrow{B} \right| .
\label{eq5_1}
\end{equation}

Consider a Lorentz transformation corresponding to the 4-velocity

\begin{equation}
  u''= (1- \frac{E^{2}}{B^{2}})^{-1/2}(1, \frac{E}{B} \hat{E} \times \hat{B})
\label{eq5_2}
\end{equation}
denoting $ \overrightarrow{E}= E. \hat{E},  \overrightarrow{B}= B. \hat{B}\qquad  (\hat{E}^{2}=1= \hat{B}^{2})$.

In the transformed frame the tensor $(\overrightarrow{E}, \overrightarrow{B})$ reduces to $(\overrightarrow{E}', \overrightarrow{B}')$ where 
\begin{equation}
  \overrightarrow{E}'=0,\qquad  \overrightarrow{B}' = (1- \frac{E^{2}}{B^{2}})^{-1/2} \overrightarrow{B} 
\label{eq5_3}
\end{equation}
such that
\begin{eqnarray}
   \overrightarrow{B}'^{2}= \overrightarrow{B}'^{2}-\overrightarrow{E}'^{2}= \overrightarrow{B}^{2}-\overrightarrow{E}^{2}\nonumber \\
   \overrightarrow{B}'. \overrightarrow{E}'=0=\overrightarrow{B}. \overrightarrow{E}.
\label{eq5_4}
\end{eqnarray}
So in this frame, one finds back the situation studied in the previous Sections II-IV, with
\begin{equation}
 B'=(B^{2}-E^{2})^{1/2}.
\label{eq5_5}
\end{equation}

The velocities and the spins of the particles are transformed according to standard rules \cite{key1,key2}. We now recapitulate some essential points.

A 4-velocity $u$ is transformed by a Lorentz transformation corresponding to  $u''$ to  $u'$ such that
\begin{equation}
  u_0'= (u_0u''_0+\overrightarrow{u}.\overrightarrow{u''}).
\label{eq5_6}
\end{equation}

Define
\begin{eqnarray}
   a &=& (1+u_0)(1+u''_0)(1+u'_0)\nonumber \\
   b &=& (1+u_0+u''_0+u'_0),
\label{eq5_7}
\end{eqnarray}
and
\begin{equation}
  \cos \frac{\delta}{2}= \frac{b}{\sqrt{2a}}.
\label{eq5_8}
\end{equation}
The spin states are conserved if $$\overrightarrow{u} \times  \overrightarrow{u''} = 0.$$ Otherwise they undergo a Wigner rotation $\delta$ about the axis

\begin{equation}
   \hat{k} = \frac{\overrightarrow{u} \times  \overrightarrow{u''} }{\left| \overrightarrow{u} \times  \overrightarrow{u''}\right| },
\label{eq5_9}
\end{equation}
such that
\begin{eqnarray}
   \left| + \right \rangle \rightarrow \left| + \right \rangle'  &=& \LP \cos \frac{\delta}{2} +i\hat{k_3} \sin \frac{\delta}{2} \RP \left| + \right \rangle \nonumber \\ &+& i\LP \hat{k_1}-i \hat{k_2}\RP \sin \frac{\delta}{2}\left| - \right \rangle \nonumber \\
   \left| - \right \rangle \rightarrow \left| - \right \rangle'  &=& i\LP \hat{k_1}+i \hat{k_2}\RP \sin \frac{\delta}{2}\left| + \right \rangle \nonumber \\ & +& \LP \cos \frac{\delta}{2} -i\hat{k_3} \sin \frac{\delta}{2} \RP \left| - \right \rangle.
\label{eq5_10}
\end{eqnarray}
The inverse rotation $(\delta \rightarrow -\delta)$ expresses $(\left| + \right \rangle , \left| - \right \rangle)$ in terms of $(\left| + \right \rangle' , \left| - \right \rangle')$.

The crucial point to note is that in \eq{eq5} apart from the $(\gamma ', \hat{v}')$ corresponding to the transformed velocities (depending on the intial velocity $\overrightarrow{v}$ and $u''$ given by \eq{eq5_2}), $B$ is replaced by  $ B'=(B^{2}-E^{2})^{1/2}$.

The magnitudes $(\omega ', \Omega ')$ thus obtained determine the modified periodicities. This is the consequence of the presence of $\overrightarrow{E}$ in the initial frame.

In the transformed frame our preceeding results (for $\overrightarrow{E}=0$) can be implemented systematically along with the velocities transformed corresponding to \eq{eq5_2}. Then the inverting of \eq{eq5_10} gives the results for the initial frame ($\overrightarrow{E} \neq 0$).

\section{Classification scheme of 3-particle entangled states}
In this Section, we propose a new classification scheme of the 3-particle entangled states, by using the eigenstates of $Z=\LP \overrightarrow{J_1} \times \overrightarrow{J_2} \RP . \overrightarrow{J_3}$ of three angular momenta.
%We briefly recapitulate some directly relevant results of Refs. \cite{key6,key7}.
One can systematically construct eigenstates \cite{key6,key7} of three coupled angular momenta ($\overrightarrow{J_1},\overrightarrow{J_2},\overrightarrow{J_3}$) by diagonalizing the operators

\begin{eqnarray}
   (\overrightarrow{J_1}+\overrightarrow{J_2}+\overrightarrow{J_3})^2\nonumber \\
   ({J_1^{(0)}}+{J_2^{(0)}}+{J_3^{(0)}})\nonumber \\
   \rm{and} \qquad Z=\LP \overrightarrow{J_1} \times \overrightarrow{J_2} \RP . \overrightarrow{J_3} ,
\label{eqa_1}
\end{eqnarray}
where $J_i^{(0)}, (i=1,2,3)$ are the projections on the z-axis.

The states are denoted by the respective eigenvalues of the above operators ($j(j+1),j^{(0)}, \zeta$) as
%\begin{equation}
 $ \left| jm \zeta \right \rangle.$
%\label{eqa_2}
%\end{equation}
Not only one obtains a complete mutually orthogonal set of eigenstates for each $j$ but along with reduction with respect to the rotation group one obtains {\it simultaneously a reduction with respect to $S_3$, the permutation group of three particles}. This is in sharp contrast with the usual 2-step reduction via 3-j coefficients where such permutations lead to 6-j coefficients.
For our purposes, we need here only the results for 
% \begin{equation}
 $ j_1=j_2=j_3=\frac{1}{2}.$
%\label{eqa_3}
%\end{equation}

In the table \ref{tab:coupling}, the states on the left correspond to eigenvalues ($j,m, \zeta$) respectively of the operators \eq{eqa_1} and those on the right to the values of $\pm\frac{1}{2}$ of ($m_1,m_2, m_3$) of $J_i^{(0)}$, ($i=1,2,3$) denoted by ($\left| 1 \right \rangle, \left| \overline{1} \right \rangle)$.
%______________________________________________________
\begin{table}[ht]
\caption{The states on the left correspond to eigenvalues ($j,m, \zeta$) respectively of the operators \eq{eqa_1} and those on the right to the values of $\pm\frac{1}{2}$ of ($m_1,m_2, m_3$) of $J_i^{(0)}$, ($i=1,2,3$) denoted by ($\left| 1 \right \rangle, \left| \overline{1} \right \rangle)$.
}
%\centering
\begin{tabular}{|l | c|}
  \hline
  $ \left| jm \zeta \right \rangle $ & $ \left| m_1 m_2 m_3 \right \rangle $ \\
    % On trace 2 lignes
  \hline
  \hline
    $ \left| \frac{3}{2} \frac{3}{2} 0 \right \rangle $ & $ \left| 111 \right \rangle $ \\
  $ \left| \frac{3}{2} \frac{-3}{2} 0 \right \rangle $ & $ \left|  \overline{1} \overline{1} \overline{1} \right \rangle $ \\
  $ \left|  \frac{3}{2} \frac{1}{2} 0 \right \rangle $ & $ \frac{1}{\sqrt3}\LP \left|   \overline{1}11 \right \rangle+\left|  1 \overline{1}1 \right \rangle+\left|   11\overline{1} \right \rangle\RP $ \\
  $ \left|   \frac{3}{2} \frac{-1}{2} 0\right \rangle $ & $ \frac{1}{\sqrt3}\LP \left|  1 \overline{1}\overline{1} \right \rangle+\left|  \overline{1}1 \overline{1} \right \rangle+\left|   \overline{1}\overline{1}1 \right \rangle \RP $ \\
  $ \left| \frac{1}{2} \frac{1}{2} \frac{\pm \sqrt3}{4} \right \rangle $ & $ \frac{1}{\sqrt3}\LP \exp({\pm i\frac{2\pi}{3}})\left|   \overline{1}11 \right \rangle+\exp({\mp i\frac{2\pi}{3}})\left|  1 \overline{1}1 \right \rangle+\left|   11\overline{1} \right \rangle \RP $ \\
  $ \left| \frac{1}{2} -\frac{1}{2} \frac{\pm \sqrt3}{4} \right \rangle $ & $ \frac{1}{\sqrt3}\LP \exp({\mp i\frac{2\pi}{3}}) \left|  1 \overline{1}\overline{1} \right \rangle+\exp({\pm i\frac{2\pi}{3}}) \left|  \overline{1}1 \overline{1} \right \rangle+\left|   \overline{1}\overline{1}1 \right \rangle\RP $ \\
  \hline
  \end{tabular}
\label{tab:coupling}
\end{table}
%\centering
% ______________________________________________________
%\begin{widetext}
%\begin{eqnarray}
%   \left| jm \zeta \right \rangle &~& \left| m_1 m_2 m_3 \right \rangle \nonumber \\
%   \left| \frac{3}{2} \frac{3}{2} 0 \right \rangle &=& \left| 111 \right \rangle \nonumber \\
%   \left| \frac{3}{2} -\frac{3}{2} 0 \right \rangle &=& \left|  \overline{1} \overline{1} \overline{1} \right \rangle \nonumber \\
%   \left|  \frac{3}{2} \frac{1}{2} 0 \right \rangle &=& \frac{1}{\sqrt3}\LP \left|   \overline{1}11 \right \rangle+\left|  1 \overline{1}1 \right \rangle+\left|   11\overline{1} \right \rangle\RP \nonumber \\
%   \left|   \frac{3}{2} -\frac{1}{2} 0\right \rangle &=& \frac{1}{\sqrt3}\LP \left|  1 \overline{1}\overline{1} \right \rangle+\left|  \overline{1}1 \overline{1} \right \rangle+\left|   \overline{1}\overline{1}1 \right \rangle \RP \nonumber \\
%   \left| \frac{1}{2} \frac{1}{2} \pm \frac{\sqrt3}{4} \right \rangle &=& \frac{1}{\sqrt3}\LP \exp({\pm i\frac{2\pi}{3}})\left|   \overline{1}11 \right \rangle+\exp({\mp i\frac{2\pi}{3}})\left|  1 \overline{1}1 \right \rangle+\left|   11\overline{1} \right \rangle \RP \nonumber \\
%   \left| \frac{1}{2} -\frac{1}{2} \pm \frac{\sqrt3}{4} \right \rangle &=& \frac{1}{\sqrt3}\LP \exp({\mp i\frac{2\pi}{3}}) \left|  1 \overline{1}\overline{1} \right \rangle+\exp({\pm i\frac{2\pi}{3}}) \left|  \overline{1}1 \overline{1} \right \rangle+\left|   \overline{1}\overline{1}1 \right \rangle\RP .
%\label{eqa_4}
%\end{eqnarray}
%\end{widetext}
This provides the complete set of 8 states spanning the space of possible values of ($m_1,m_2, m_3$).
Thus 
\begin{equation}
  \frac{1}{\sqrt2}\LP \left| \frac{3}{2} \frac{3}{2} 0 \right \rangle\pm \left| \frac{3}{2} -\frac{3}{2} 0 \right \rangle \RP=\frac{1}{\sqrt2}\LP \left| 111 \right \rangle \pm \left|  \overline{1} \overline{1} \overline{1} \right \rangle\RP,
\label{eqa_5}
\end{equation}
giving the $\left| GHZ \right \rangle$ states as a doublet.

The others correspond directly to the Werner ($\left| W \right \rangle$) and the flipped Werner ($\left| \widetilde{W} \right \rangle$) and their variants with relative phases $\exp({\pm i\frac{2\pi}{3}})$, which we had mentioned in the remarks below \eq{eq6} in Section II.

Of the three operators in  \eq{eqa_1} the first two are invariant under all permutations of the particles ($1,2,3$). The remaining one ($Z$) is invariant under circular permutations of ($1,2,3$) and just changes sign under ($12$),($23$) and ($31$). This is in sharp contrast to the standard 2-step couplings where one has to pass from one 3-j coupling scheme to another under permutations. This is at the root of the simultaneous reduction under $S_3$ via the implementation of $Z$. This also helps to explain the direct relations of the $Z$-eigenstates with 3-tangles invariant under  permutations of ($1,2,3$).

\section{Concluding remarks and outlook}
We summarize the essential features of the present work, and give an outlook of future directions of work:
\begin{enumerate}
\item An external (constant) magnetic field induces a precession of the spin of each particle of the entangled states considered, through local unitary transformations. Note that for two particle states of total spin zero, studied in earlier paper, the periodic precession of individual spins was not present. Here we analysed three particle states where such \textit{periodicities} (2 and 4 periods respectively for the particular cases illustrated) in correlations and density matrices were remarkably displayed and intertwined. The patterns of periodicity thus emerging were shown to be remarkably rich and subtle. Such a study was presented for the first time.

\item The individual precessions being implemented by local unitary matrices, the initial 3-tangle was conserved. But simply the verification of this fact was not the aim of the analyses of sections III and IV. We went beyond that and displayed in details how the conservation left scope for component periodic correlations to appear, increase and decrease. We pinpointed the crucial roles of the signs of the coefficients of different components, again for the first time.

\item The generalization to include orthogonal electric field was presented in Section V, using the Wigner rotation. For $E=B$, the Lorentz transformation via \eq{eq5_2} is not well-defined and for $E>B$ it becomes complex. The limiting case $E=B$ (e.g. for a plane wave field) will be studied elsewhere using exact solutions \cite{key10} of the Dirac-Pauli equations in such fields. 

\item The remarkable and systematic correspondence of famous entangled states to a specific coupling scheme for 3-angular momenta was presented in Section VI, and a new classification scheme was proposed. We intend to study this aspect in details elsewhere.

\item Finally one may note that unitary transformations may be {\it non-local} when induced via unitary braid matrices \cite{key11}. Acting on pure product states, they can then {\it generate} entanglement. In the present work we started with states already entangled and then followed their periodic ramifications as they evolved in the magnetic fields, which is completely different.
\end{enumerate}

%\begin{acknowledgments}
%A.C. is very grateful to P.B. Pal
%for stimulating discussions and useful criticism.
%\end{acknowledgments}

\end{document}